\documentclass[a4paper]{jpconf}
\usepackage{graphicx,amsmath,amsfonts,bm}
\begin{document}
\title{Reconstruction algorithm in compressed sensing based on maximum a
posteriori estimation}

\author{Koujin Takeda$^{1}$ and Yoshiyuki Kabashima$^{2}$}

\address{Department of Intelligent Systems Engineering, Ibaraki University$^{1}$ \\
Department of Computational Intelligence and
Systems Science, Tokyo Institute of Technology$^{2}$}

\ead{ktakeda@mx.ibaraki.ac.jp}

\begin{abstract}
 We propose a systematic method for constructing a sparse data reconstruction algorithm 
 in compressed sensing at a relatively low computational cost for general
 observation matrix.
 It is known that the cost of $\ell_1$-norm minimization using a standard linear
 programming algorithm is $O(N^3)$.
 We show that this cost can be reduced to $O(N^2)$
 by applying the approach of posterior maximization.
 Furthermore, in principle, the algorithm from our approach is 
 expected to achieve the widest successful reconstruction region, which is evaluated 
 from theoretical argument.
 We also discuss the relation between the belief propagation-based
 reconstruction algorithm introduced in preceding works and our approach.
\end{abstract}

\section{Introduction}
Nowadays, use of the compressed sensing (CS) approach \cite{Donoho,
CandesTao, CRT} is rapidly
spreading to various fields in information technology, where the
sparsity of the original data plays a crucial role
\cite{CSreview}.

In this article, we present our study on a very basic problem of CS.
Let us consider a linear observation process expressed as
\begin{equation}
 \bm y = \bm F \bm x,
\end{equation}
where $\bm y \in \mathbb{R}^M$ is the observed data
 and $\bm x \in \mathbb{R}^N$ is the original sparse data with
many zero entries. The (averaged) number of nonzero entries is
denoted by $K$. The $M \times N$ matrix $\bm F$ describes the
process of observation, and the limit of large $M,N$ with
compression rate $\alpha := M/N<1$ is taken into account. (Throughout this
paper, bold symbols denote a vector/matrix.)

We investigate the basic $\ell_1$-norm minimization
\begin{equation}
\label{eq:ell1}
{\rm min}_{\bm x}\| \bm x \|_{1}
 \ \ {\rm subject\ to} \ \ \bm y = \bm F \bm x.
\end{equation}
This problem is written as linear programming and can be solved at
a computational cost of $O(N^3)$ using a standard algorithm, such
as the interior point method. However, many alternative algorithms
have been proposed for reducing the computational cost
\cite{Elad,TW}. Among these, Approximate Message Passing (AMP)
\cite{DMM}, which is a thresholding algorithm that is based on the
message passing approach \cite{Pearl}, is noteworthy. In this
algorithm, each entry of the matrix $\bm F$ is assumed to be
random with identical Gaussian distribution. Then, it is shown
that the original data can be reconstructed at a computational cost
of $O(N^2)$. Furthermore, this algorithm has theoretical
significance: for the AMP, the successful reconstruction threshold in
terms of compression rate $\alpha$ and data sparsity $\rho := K/N$ is
analytically shown to be the same as for $\ell_1$-norm
minimization \cite{DT} using a state evolution technique \cite{DMM,DMM3,DMM4,BM}.
However, it is not intuitively evident from their argument
why the threshold of the AMP agrees with that of $\ell_1$-norm
minimization.

In this article, we propose an approach for constructing a sparse
data reconstruction algorithm by using maximum {\it a posteriori}
probability (MAP) for general matrix $\bm F$
with as small an $M$ as possible. Using this
approach, we obtain an algorithm whose computational cost is
$O(N^2)$. (For a special case, when the matrix $\bm F$ is sparse
and has only $O(1)$ nonzero entries in each column/row, the
computational cost is reduced to $O(N)$.) In addition, our
approach explains from another perspective why the reconstruction
thresholds of the AMP and $\ell_1$-norm minimization are analytically
equivalent.

\section{Reconstruction algorithm using the MAP}
To construct our MAP algorithm, first, we prepare the posterior
probability. We need to infer the original data $\bm x$ from the
observed data $\bm y$ and the matrix $\bm F$; therefore, we define
the posterior probability $P (\bm x | \bm y, \bm F)$ for $\bm x$
 in accordance with the $\ell_1$-norm minimization problem as
(\ref{eq:ell1})
\begin{eqnarray}
\label{eq:Pdist}
 P(\bm x | \bm y, \bm F) 
 &:=& \lim_{\beta \rightarrow \infty}
 \! \frac{1}{Z(\beta)}  \exp \left( \! - \beta  \sum_i | x_i | \! \right)
 \prod_{\mu} \delta \left( \!
 y_{\mu} - \sum_i F_{\mu i} x_i \! \right)\!,
\end{eqnarray}
where $Z(\beta)$ is the normalization,
\begin{eqnarray}
Z(\beta)&:=& \exp(-\beta C), \nonumber \\
C&:=& {\rm min}_{\bm x} \| \bm x \|_{1}
 \ \ {\rm subject\ to} \ \ \bm y = \bm F \bm x.
\end{eqnarray}
(Roman subscripts run from 1 to $N$, and Greek from 1 to $M$).
This probability is zero unless the constraint $\bm y = \bm F \bm
x$ is satisfied. In addition, it is unity only if $\ell_1$-norm of
$\bm x$ is minimum; otherwise zero. Then, it is found that the MAP
solution is equivalent to the $\ell_1$-norm minimum. Note that we
do not need to take the limit of $\beta \rightarrow \infty$ for a
MAP solution; however, we take it for clarifying the relation
between the MAP algorithm and the reconstruction threshold
analysis using a statistical mechanical replica method \cite{KWT}
that gives the exact expression of the threshold \cite{DT}, where
this posterior probability is defined as the Boltzmann weight, and
zero temperature limit $\beta \rightarrow \infty$ is taken for
technical reasons related to the analysis.

An obstacle to dealing with posterior probability is the singular
delta function; we regularize this by the quadratic term
\begin{eqnarray}
\label{eq:Pdist}
P(\bm x | \bm y, \bm F) &=&
 \lim_{\beta \rightarrow \infty} \! \frac{1}{Z (\beta)}
\exp \left(\!\!
 - \beta \frac{\sum_{\mu}
 (y_{\mu} - \sum_i F_{\mu i} x_i)^2}{2}
 - \! \beta k \sum_i | x_i | \! \right).
\end{eqnarray}
This posterior reproduces the original $\ell_1$ minimization
problem by taking the limit $\beta \rightarrow \infty$. 
(The readers can find a similar form of the probability density
in \cite{DMM3} through the discussion of the AMP algorithm.)
In this formulation, we need to introduce a constant $k(>0)$ that
represents the ``relative significance'' of constraint and
minimization. For $\ell_1$-norm minimum solution, we must consider
the limit $k \rightarrow 0$ because we are attempting to find the
minimum solution {\it under} the constraint at present. This framework
is essentially the same as that used in the lasso \cite{lasso};
however, the crucial point here is that we need to take the limit
$k \rightarrow 0$ for the MAP solution appropriately.

For the MAP solution, we differentiate the term in the exponential
with respect to $x_i$, and write the stationary condition
\begin{eqnarray}
 \sum_{\mu} F_{\mu i} y_{\mu}
\! - \! \sum_{j (\ne i)} \! \sum_{\mu} F_{\mu i} F_{\mu j} x_j
\! - \! \sum_{\mu} F_{\mu i}^2 x_i
\! - \! k \Theta(x_i) = 0
\end{eqnarray}
(where $\Theta(x)$ is the Heaviside function), which can be
rewritten as
\begin{eqnarray}
\label{eq:MAPz}
x_i &=& \frac{1}{ \sum_{\mu} F_{\mu i}^2} \ \
\eta \left(  \sum_{\mu} F_{\mu i}
\left( z_{\mu} + F_{\mu i} x_i \right) ; k
\right), \nonumber \\
z_{\mu} &:=& y_{\mu} - \sum_{i} F_{\mu i} x_i.
\end{eqnarray}
Here, for convenience, we define the function for thresholding
\begin{eqnarray}
\label{eq:thresfunc}
\eta(x;k) := \left\{
\begin{array}{cc}
x - k & k < x \\
0     & -k \le x \le k \\
x + k & x < -k
\end{array} \right. ,
\end{eqnarray}
and also introduce the variable $z_{\mu}$, which represents the
residual error of the constraint $\bm y = \bm F \bm x$. Basically,
we can obtain the $\ell_1$-norm minimum solution by solving this
stationary condition. We construct an iterative algorithm by
adding the iteration step superscript $(t)$:
\begin{eqnarray}
\label{eq:naivealgo}
x_i^{(t)} &=& \frac{1}{ \sum_{\mu} F_{\mu i}^2} \ \
\eta \left(  \sum_{\mu} F_{\mu i}
\left( z_{\mu}^{(t)} + F_{\mu i} x_i^{(t-1)} \right) ; k
\right), \nonumber \\
z_{\mu}^{(t)} &=& y_{\mu} - \sum_{i} F_{\mu i} x_i^{(t-1)}.
\end{eqnarray}
The MAP solution is obtained by finding the fixed point of
$x_i^{(t)}$. Remember that the original $\ell_1$-norm minimization
is a convex optimization, and accordingly there is no fixed point
of local minimum.

We now give some remarks about this algorithm (\ref{eq:naivealgo}).
\begin{itemize}
\item{Computational cost}

In algorithmic equations (\ref{eq:naivealgo}), only a single
summation appears, and the numbers of the equations for
$x_i^{(t)}$ and $z_{\mu}^{(t)}$ are $N, M$, respectively. Then,
the computational cost is $O(N^2)$ when the compression rate $M/N$
is $O(1)$. In addition, if the matrix $\bm F$ is sparse (having
only $O(1)$ nonzero entries in each column/row, which
is discussed in \cite{KW, KMSSZ, KMSSZ2}), the computational
cost is reduced to $O(N)$. Note that the number of iteration steps
for convergence is assumed to be much smaller than the orders of
$M$ and $N$. However, in general, it does not hold near the region
of the reconstruction threshold, where the speed of
$k \rightarrow 0$ limit must be slow for convergence and
the number of iteration steps for $k \rightarrow 0$ 
becomes dominant in comparison with the orders of $M$ and $N$.

\item{Applicability to general $\bm F$}

We do not make any assumption for $\bm F$. Therefore, in
principle, this algorithm can be used for general matrix $\bm
F$, whereas in the original AMP \cite{DMM} or in the statistical
mechanical analysis \cite{KWT} i.i.d. random entries are assumed.

\item{$k \rightarrow 0$ limit}

We must take the $k \rightarrow 0$ limit to the final iteration
step. Empirically, $k$ should be decreased exponentially with the
step as $k^{(t)} \propto \exp (- t / {\rm const.})$. The decay
constant in the exponential is significant because too small a
constant leads to a wrong solution. This constant is found to be
the threshold parameter in the AMP algorithm equation as will be
elucidated later, which is significant for convergence as stated in \cite{DMM}.
In the original work of the AMP \cite{DMM}
the authors discussed the appropriate
limit of $k \rightarrow 0$ by choosing $k$ as the mean squared
error, and, in conjunction with the state evolution technique,
they arrive at the reconstruction threshold
for convergence, which is represented by the relation between $\alpha$ and $\rho$
with auxiliary variable $z$, 
\begin{eqnarray}
\label{eq:AMPthreshold}
\rho = \alpha\ \underset{z \ge 0}{\rm max} 
\left( \frac{1 - \frac{2}{\alpha} \left\{ (1+z^2) H(z) 
- z \frac{e^{-\frac{z^2}{2}}}{\sqrt{2\pi}} \right\} }
{1+z^2 - 2 \left\{ (1+z^2) H(z) - z \frac{e^{-\frac{z^2}{2}}}{\sqrt{2\pi}} \right\} }
\right),
\end{eqnarray}
where $H(z):= \int_z^{+\infty} dx e^{-x^2/2} / \sqrt{2\pi}$.
This expression is $\it analytically$ equivalent to that of $\ell_1$-norm minimization
evaluated by combinatorial geometry \cite{DT,DT2}.  
Replica method can also give the threshold equations \cite{KWT} as
\begin{eqnarray}
\label{eq:replicathreshold}
2(1-\rho) \left( H(z) - \frac{1}{z} \frac{e^{-\frac{z^2}{2}}}{\sqrt{2\pi}} \right)
+ \rho = 0, \nonumber \\
\alpha = 2(1-\rho) H(z) + \rho,
\end{eqnarray}
which is also analytically equivalent. (Readers can check the equivalence between (\ref
{eq:AMPthreshold}) and (\ref{eq:replicathreshold}) after some algebra.)

As elucidated above, our initial step for the algorithm is
the probability distribution in (\ref{eq:Pdist}), which appears as
Boltzmann weight in replica analysis for computation of the threshold. 
In the AMP, the convergence condition (\ref{eq:replicathreshold}) 
is discussed using the thresholding algorithm {\it without} Onsager term 
(namely, from the na\"{i}ve algorithm (\ref{eq:naivealgo})), which
also originates from the distribution (\ref{eq:Pdist}).
This relation explains why we arrive at the same threshold equation.

\item{Partial update}

As is widely recognized, the na\"{i}ve thresholding algorithm 
(\ref{eq:naivealgo}) still presents a
problem. We cannot always obtain a correct $\ell_1$-norm minimum
solution using (\ref{eq:naivealgo}), even in successful reconstruction
region from (\ref{eq:AMPthreshold}, \ref{eq:replicathreshold})
and even when the limit $k\rightarrow 0$ is suitably designed. 
Such a failure occurs near
the reconstruction threshold of $\ell_1$-norm minimization. (Readers 
will find the details below.) The AMP successfully solves
this problem by the introduction of Onsager reaction term,
which is reduced from perturbative analysis in the original work \cite{DMM}. 
We attempt to
address this problem from another point of view, which is applicable
for general matrix $\bm F$.

The strategy is as follows. 
In the original algorithm, we fully update the variable $x_i^{(t)}$ in each step.
Now, we introduce a partition ratio in the algorithm, and modify
the algorithm to include the partial update rule as
\begin{eqnarray}
\label{eq:partitionalgo}
x_i^{(t)} &=& \frac{\gamma^{(t)}}{1+\gamma^{(t)}} x_i^{(t)}
+ \frac{1}{1+\gamma^{(t)}}
 \frac{1}{ \sum_{\mu} F_{\mu i}^2} \,\,
\eta \! \left(  \sum_{\mu} F_{\mu i}
\! \left( z_{\mu}^{(t)} + F_{\mu i} x_i^{(t-1)} \right) \! ; k
\! \right)\!, \nonumber \\
z_{\mu}^{(t)} &=& y_{\mu} - \sum_{i} F_{\mu i} x_i^{(t-1)},
\end{eqnarray}
where the partition ratio $\gamma^{(t)}\ $ is dependent on the
step in general. Obviously, the fixed point of the algorithm is
the same as that of the original; nevertheless, the convergence to
a correct solution is improved by choosing $\gamma^{(t)}$
properly.

Here $\gamma^{(t)}$ can be chosen arbitrarily. We can discuss 
how to design $\gamma^{(t)}$ to achieve better
convergence by the stability analysis around the fixed point of
the algorithm. However, at present we do not have the best prescription
of how to choose $\gamma^{(t)}$. 
We omit the details of the discussion here due to limited space.
\end{itemize}

\section{Discussion on the MAP algorithm}

Let us move on to the properties of the proposed algorithm. Here,
we use the matrix $\bm F$ with i.i.d random entries drawn from
Gaussian distribution with zero mean and variance $1/M$, which is
the same as in the AMP \cite{DMM}
and statistical mechanical analysis \cite{KWT}. For large
$M,N$, we have $\lim_{M \rightarrow \infty} \sum_{\mu} F_{\mu i}^2
= 1$ and the algorithm equation (\ref{eq:naivealgo}) is simplified
to
\begin{eqnarray}
\label{eq:algoorig}
x_i^{(t)} &=& \eta \left(  \sum_{\mu} F_{\mu i}  z_{\mu}^{(t)}
 +  x_i^{(t-1)} ; k \right), \nonumber \\
z_{\mu}^{(t)} &=& y_{\mu} - \sum_{i} F_{\mu i} x_i^{(t-1)},
\end{eqnarray}
which can also be found in the introductory part of \cite{DMM},
where it is mentioned that the performance of this na\"{i}ve
thresholding algorithm is worse than that of $\ell_1$-norm
minimization. We introduce a partition ratio in the first line of
(\ref{eq:algoorig}) (here $\gamma$ is not dependent on the step
$t$) to improve it:
\begin{eqnarray}
\label{eq:partition}
x_i^{(t)} \!&=&\! \frac{\gamma}{1+\gamma} x_i^{(t-1)}
 + \frac{1}{1+\gamma}
 \eta \left(  \sum_{\mu} F_{\mu i}  z_{\mu}^{(t)} +  x_i^{(t-1)} ; k
\right).
\end{eqnarray}
We rewrite this equation as
\begin{eqnarray}
\label{eq:partitionupdate}
x_i^{(t)} \!\!\!\! &=& \!\!\!\! \frac{\gamma}{1+\gamma} x_i^{(t-1)} +
 \! \eta \! \left( \frac{1}{1+\gamma} \! \left\{ \sum_{\mu} F_{\mu i}
  z_{\mu}^{(t)} +  x_i^{(t-1)} \right\} ; k \! \right)\!,
\end{eqnarray}
using the property of $\eta$. (We rescale the parameter $k$, which
is irrelevant to the fixed point because $k$ should be taken to
zero, finally.) Instead of (\ref{eq:partitionupdate}), we now introduce
slightly modified update rule
\begin{eqnarray}
\label{eq:partitionupdate2}
x_i^{(t)} &=&  \eta \left( \frac{1}{1+\gamma}
 \sum_{\mu} F_{\mu i}  z_{\mu}^{(t)} +  x_i^{(t-1)} ; k \right)
\end{eqnarray}
by moving the first term on r.h.s. into the argument of $\eta$.
This gives the same fixed point as (\ref{eq:partitionupdate}), which can be verified 
by eliminating the step superscript (t) from both update rules (\ref{eq:partitionupdate},
\ref{eq:partitionupdate2}) 
and solving them with respect to $x$.
With the rescaled residual error $\widehat{z}_{\mu}^{(t)}
 =  z_{\mu}^{(t)} / (1+\gamma)$, we obtain
\begin{eqnarray}
x_i^{(t)}
\! &=& \! \eta \left( \sum_{\mu} F_{\mu i} \widehat{z}_{\mu}^{(t)}
 + x_i^{(t-1)} ; k \right),
 \nonumber \\
\widehat{z}_{\mu}^{(t)}
 &=& \frac{1}{1+\gamma} \left\{ y_{\mu}
 - \sum_{i} F_{\mu i} x_i^{(t-1)} \right\}.
\end{eqnarray}
This expression of the algorithm indicates that the partition
ratio is associated with the scaling of the residual error
$z_{\mu}$.

Next, we consider the step-dependent $\gamma^{(t)}$. In
particular, we choose
\begin{eqnarray}
\label{eq:gammavary}
\gamma^{(t)} &=& \frac{1}{M}
 \sum_i \eta' \left( \sum_{\mu} F_{\mu i} z_{\mu}^{(t)}
 + x_i^{(t-1)} ; k \right),
\end{eqnarray}
where $\eta'(x;k) := \partial_x \eta(x;k)$.
This $\gamma^{(t)}$ is chosen in order to achieve faster convergence 
according to the second order stability analysis around the solution,
however the details are omitted. The point is that $\gamma^{(t)}$
is expressed by the function for thresholding, namely equation (\ref{eq:thresfunc}), which also
appears in the algorithm. Note that $\gamma^{(t)}$ approaches the
value $K/M = \rho/\alpha$ for $t \rightarrow \infty$ when the reconstruction is
successful. For this step-dependent $\gamma^{(t)}$, we can also
reach another expression of the algorithm in a similar manner as
constant $\gamma$, 
by using that $\gamma^{(t)}$ is represented by the function (\ref{eq:thresfunc}),
\begin{eqnarray}
\label{eq:partitionfinal}
x_i^{(t)} &=&  \eta \left(
 \sum_{\mu} F_{\mu i}  \widehat{z}_{\mu}^{(t)} +  x_i^{(t-1)} ; k
            \right), \nonumber \\
\widehat{z}_{\mu}^{(t)} &=&
  \frac{1}{1+ \frac{1}{M}
  \sum_j \eta' \left( \sum_{\mu} F_{\mu j} z_{\mu}^{(t)}
  + x_j^{(t-1)} ; k \right)} z_{\mu}^{(t)},
 \nonumber \\
z_{\mu}^{(t)} &=& y_{\mu} - \sum_{i} F_{\mu i} x_i^{(t-1)}.
\end{eqnarray}
This expression indicates that the scaling of the residual error $z_{\mu}$ varies
as a function of the step $t$.

\section{Relation to the AMP}
Based on the observation above, we choose the step-dependent
partition ratio $\gamma^{(t)}$ with negative sign
\begin{eqnarray}
\gamma^{(t)} = - \frac{1}{M}
 \sum_i \eta'
 \left( \sum_{\mu} F_{\mu i} z_{\mu}^{(t)} + x_i^{(t-1)} ; k \right),
\end{eqnarray}
which leads to partial update rule by external division.
In this case we have almost the same equations as in (\ref{eq:partitionfinal}); 
the only difference is the negative sign in the denominator in the second equation.
 Here, we also introduce the partition ratio
$\widehat{\gamma}^{(t)}$ for the update rule of the 
rescaled residual error $\widehat{z}_{\mu}^{(t)}$,
\begin{eqnarray}
\label{eq:zpartition}
\widehat{z}_{\mu}^{(t)} &=& \widehat{\gamma}^{(t-1)} \widehat{z}_{\mu}^{(t-1)}
 +  \frac{1 - \widehat{\gamma}^{(t-1)}}{1- \frac{1}{M}
  \sum_j \eta'
 \left( \sum_{\mu} F_{\mu j} z_{\mu}^{(t)} + x_j^{(t-1)} ; k \right)}
 z_{\mu}^{(t)}. 
\end{eqnarray}
By choosing $\widehat{\gamma}^{(t)} = - \gamma^{(t)} =
\sum_{j} \eta'(\sum_{\mu} F_{\mu j} z_{\mu}^{(t)}
 + x_j^{(t-1)} ) / M$, the algorithm is changed to
\begin{eqnarray}
 x_i^{(t)} &=&  \eta \left(
 \sum_{\mu} F_{\mu i}  \widehat{z}_{\mu}^{(t)} +  x_i^{(x-1)} ; k
 \right), \nonumber \\
\widehat{z}_{\mu}^{(t)}\! &=& \!
  {z}_{\mu}^{(t)}
 + \frac{1}{M} \widehat{z}_{\mu}^{(t-1)}
   \sum_j \eta' \left(
 \sum_{\mu} F_{\mu j} z_{\mu}^{(t)} + x_j^{(t-1)} ; k \right),
 \nonumber \\
z_{\mu} &=& y_{\mu} - \sum_{i} F_{\mu i} x_i.
\end{eqnarray}
We arrive at this expression just by introducing
the partition ratios and rewriting the update rule. When we replace $z_{\mu}^{(t)}$ with
$\widehat{z}_{\mu}^{(t-1)}$ in the second equation above
(approximation by $z \rightarrow \widehat{z}$, which does not
change the fixed point because $z$ and $\widehat{z}$ finally
approach zero), and shift the step number of $x_i^{(t-1)}$ in
$\eta'$ to $t-2$ (which does not change the fixed point either),
we finally obtain
\begin{eqnarray}
\label{eq:AMP}
 x_i^{(t)} &=&  \eta \left(
 \sum_{\mu} F_{\mu i}  \widehat{z}_{\mu}^{(t)} +  x_i^{(t-1)}
 ; k \right), \nonumber \\
\widehat{z}_{\mu}^{(t)} &=&
   y_{\mu} - \sum_{i} F_{\mu i} x_i^{(t-1)}
 + \frac{1}{M} \widehat{z}_{\mu}^{(t-1)}
  \sum_j \eta' \left( \sum_{\mu} F_{\mu j}
 \widehat{z}_{\mu}^{(t-1)} + x_j^{(t-2)} ; k \right),
\end{eqnarray}
which is nothing but the update rule of the variables in the AMP
\cite{DMM}, where the parameter $k$ controls the thresholding.
Thus, we can find the relation between the AMP and the partial update
MAP algorithm. In the AMP, the last term in the second
equation in (\ref{eq:AMP}) is derived from the discussion of
message passing, and is interpreted as the Onsager reaction term
in statistical mechanics \cite{DMM,BM}.
Using the logic here, we can provide a
viewpoint that this term is introduced by way of a partition ratio
in order to achieve better convergence of the algorithm.

\section{Numerical experiment}

\subsection{Reconstruction threshold}
First, we evaluated the reconstruction threshold of our algorithm.
In the experiment, the original data dimension was fixed as
$N=10^3$. The numbers of observations $M$ and (averaged) nonzero
data components $K$ were varied. Each entry of $\bm F$ was drawn
randomly and independently from Gaussian distribution with zero
mean and variance $1/M$ (in Figures 1-3). Each nonzero element of $\bm x^0$ was
similarly drawn from Gaussian distribution with zero mean and unit
variance. The initial value of $x_i^{(t)}$ was set to be zero. The
parameter $k$ approached zero in each step exponentially by
multiplying the factor $0.999$ in each update, that is, a very slow approach to
the limit of $k \rightarrow 0$ was used to guarantee convergence.
We took the value $x_i^{(t)}$ as the reconstruction result after
$5 \times 10^3$ steps in Figures 1 and 2, and $1 \times 10^4$ in
Figures 3 and 4, respectively. For evaluating the reconstruction
threshold, we conducted the experiment 50 times and computed the
success rate for fixed $M,K$. In each trial, the reconstruction
was judged to be a success if the mean squared error per data was
less than $10^{-3}$. Then, we computed the success rate for every
pair of $M,K$ by changing their values. (After the experiment
for one pair of $M,K$, we increased/decreased $M$ or $K$ by 25.)

The results are shown in Figures 1-4. The success rate obtained in
the experiment is expressed by gray-scale in each figure. In
Figure 1, the result by the algorithm without
partition ratio is shown, which exhibits narrower success region
than that of $\ell_1$-norm minimization. (This threshold 
was also evaluated in \cite{DMM}.) 
The result with the partition ratio $\gamma^{(t)}=1$
is shown in Figure 2, where a reconstruction threshold close to
$\ell_1$-norm threshold curve is observed. Then, for improvement
we used the step-dependent partition ratio in (\ref{eq:gammavary}). 
With this modification, we can obtain much closer threshold 
to $\ell_1$-norm curve as in Figure 3. In lower compression rate
(=large $\alpha$) region, the performance seems
slightly worse than that of $\ell_1$-norm minimization. However,
we expect that larger number of iteration steps, appropriate
choice of $k \rightarrow 0$ limit, and suitable partition ratio
$\gamma^{(t)}$ will improve the performance. Next, for checking 
the applicability to general matrix $\bm F$, we study
the case where the random matrix $\bm F$ is generated by a
different rule: we first generate a dense random matrix $\bm F$ 
using the same rule as before, and then randomly eliminate 90\% of the
entries, which are set to be zero. In this experiment we use the algorithm 
(\ref{eq:partitionalgo}) with the partition ratio $\gamma^{(t)}=1$. 
The result in Figure 4 exhibits a success region similar to 
that of $\ell_1$-norm, which can be understood from reconstruction 
threshold universality for wide class of random $\bm F$, 
as indicated in \cite{DT3,DMM} and theoretically discussed in \cite{KWT, BLM, TK}.

From these results, we confirm that the MAP algorithm is
essentially equivalent to $\ell_1$-norm minimization, and through
a suitable design of the algorithm, we can expect a performance
that is almost the same
 as that of $\ell_1$-norm. We  stress that the
algorithm proposed here will also be successful under the general
matrix $\bm F$, as observed in Figure 4.

\begin{figure}
\begin{picture}(100,160)
\put(10,-5){\includegraphics[width=0.60\textwidth]{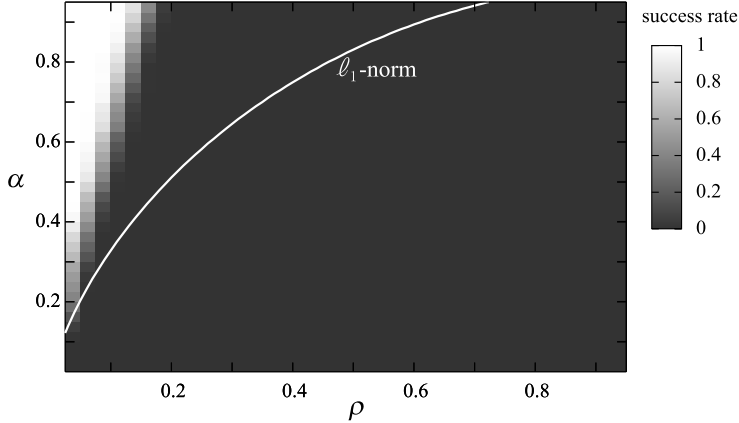}}
\end{picture}
\caption{Profile of the success rate
 using the algorithm without partition ratio (\ref{eq:naivealgo}).
 Success rate is displayed by gray-scale.
 As seen, the success region is not as wide
 as with $\ell_1$-norm minimization.}
\label{fig1}
\end{figure}

\begin{figure}
\begin{picture}(100,160)
\put(10,-5){\includegraphics[width=0.60\textwidth]{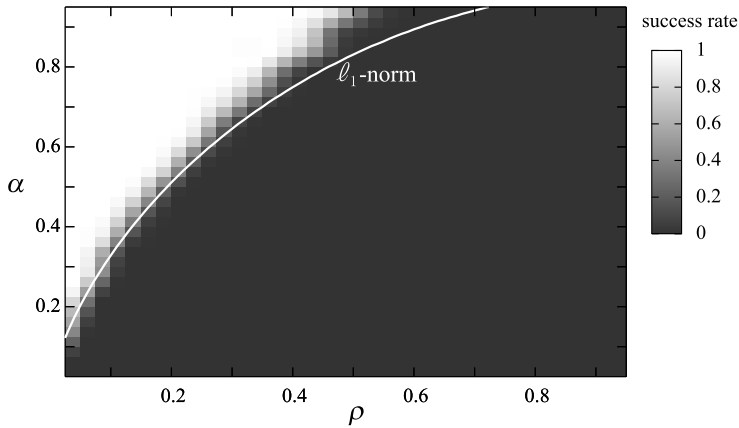}}
\end{picture}
\caption{Profile of the success rate using the algorithm
(\ref{eq:partitionalgo}) with the partition ratio
$\gamma^{(t)}=1$. The area of the success region is almost the
same as that of $\ell_1$-norm minimization.} \label{fig2}
\end{figure}

\begin{figure}
\begin{picture}(100,160)
\put(10,-5){\includegraphics[width=0.60\textwidth]{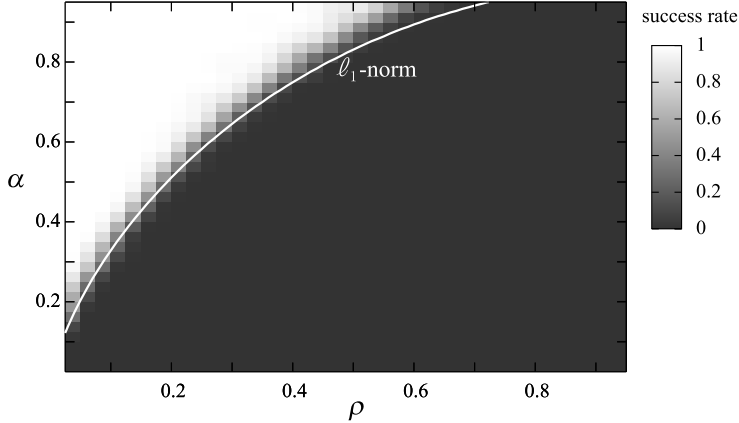}}
\end{picture}
\caption{Profile of the success rate using the algorithm with
step-dependent partition ratio $\gamma^{(t)}$ in
(\ref{eq:gammavary}). We can obtain a  reconstruction threshold
that is much closer to that of $\ell_1$-norm minimization.}
\label{fig3}
\end{figure}

\begin{figure}
\begin{picture}(100,160)
\put(10,-5){\includegraphics[width=0.60\textwidth]{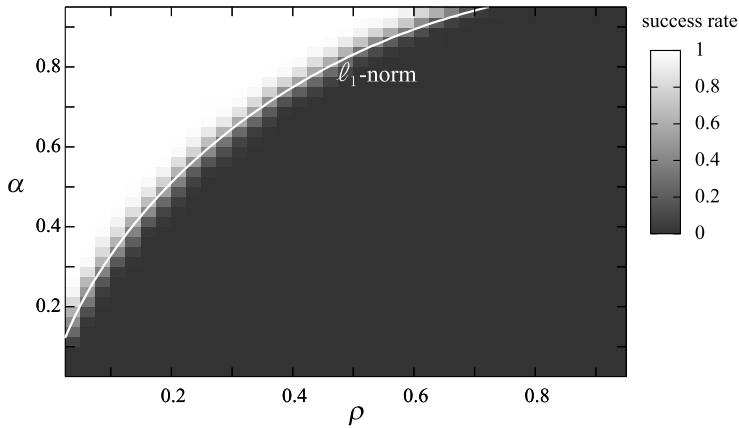}}
\end{picture}
\caption{Profile of the success rate for a random matrix with a
10\% nonzero matrix element. We use the algorithm with the
constant partition ratio $\gamma^{(t)}=1$. We can obtain a
success region similar to that of $\ell_1$-norm minimization also
in this case.} \label{fig4}
\end{figure}

\subsection{Convergence of the algorithm}

Next, we examined the speed of convergence to the solution. We
compared two algorithms: the MAP algorithm with step-dependent
partition ratio (\ref{eq:gammavary}), and the AMP (\ref{eq:AMP}).
We set the parameters $N=2000, M=1000$ (compression ratio $\alpha=0.5$)
and $K=200$ (fraction of nonzero entries $\rho=0.1$). The rules for
generating the matrix $\bm F$ and the original data $\bm x^0$ were
the same as in Figures 1-3. For the parameter $k$, the exponential
decay limit to zero was taken by multiplying $0.95$ in each step for
both algorithms. (Note that the original AMP uses the mean squared error
for the update of $k$ as elucidated before, which shows much better 
convergence. Here we adopt slow exponential update rule for both algorithms
in the experiment for simpler experimental condition.)
We conducted the reconstruction experiment 100 times and observed
the behavior of the mean square error per data. The results are
shown in Figure 5. It can be seen that the AMP exhibits a better
performance than the MAP algorithm. From the relation between the
MAP and the AMP algorithms, as discussed in the previous section,
this difference in convergence speed might be understood from the
introduction of the partition ratio to the residual error $z_{\mu}$ to
improve the convergence in the AMP (as in ({\ref{eq:zpartition})),
whereas in the MAP algorithm in (\ref{eq:partitionalgo}) this
point is not taken into consideration. Therefore, it will be
necessary to consider faster convergence of the residual error
$z_{\mu}$ to improve the MAP algorithm.

\begin{figure}
\begin{picture}(100,150)
\put(20,-8){\includegraphics[width=0.51\textwidth,angle=0]{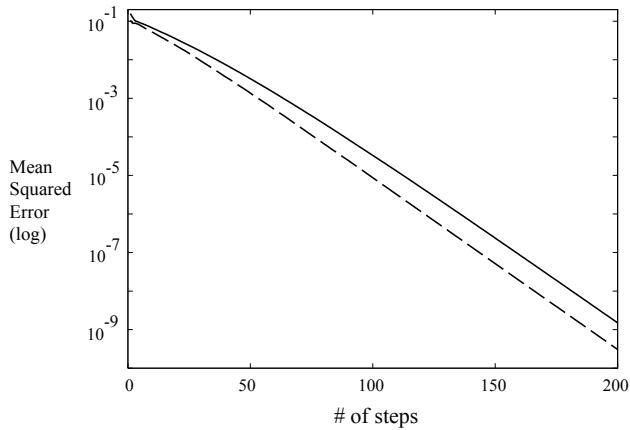}}
\end{picture}
\caption{Convergence speed of the algorithm. We compare the
convergence speed of the algorithms,
 the MAP with a step-dependent partition ratio (solid)
 and the AMP (broken).
The behavior of the mean squared error per single data point is
depicted. The AMP exhibits a better performance than the MAP in terms of
convergence speed.} \label{fig4}
\end{figure}

\section{Conclusion and perspective}
In this article, we presented a methodology for constructing an
algorithm using the MAP approach, discussed its relation with a
known thresholding algorithm, and evaluated its performance
through numerical experiments. We verified that, by  designing the
algorithm appropriately, almost the same reconstruction threshold
as that of $\ell_1$-norm minimization can be achieved. In the case
of the i.i.d. random matrix $\bm F$, we clearly presented a
viewpoint on the reason why the AMP has the same analytical expression
of the reconstruction threshold as $\ell_1$-norm minimization.
The significant point is that we do not make any assumption for
the matrix $\bm F$, and accordingly our algorithm's construction,
whose computational cost is relatively low, is applicable to a
general matrix.

We also emphasized that it is still possible that faster
convergence to the correct solution can be achieved by a more
suitable design of the partition ratio. In the case of the random
i.i.d. Gaussian matrix, the optimality of the algorithm is
discussed in \cite{BM}. For a general matrix such as sparse or 
structured matrix, for which feasible algorithm is proposed and
discussed in \cite{KW,KMSSZ,KMSSZ2}, future work will address 
a systematic method for designing an
algorithm that will achieve faster convergence
using the discussion presented in this paper.

\section*{Acknowledgments}

This work is supported by KAKENHI
Nos. 24700007 (KT), 22300003 and 25120013 (YK).
YK also acknowledges the ELC project (Grant-in-Aid for
Scientific Research on Innovative Areas MEXT Japan) for encouraging
the research presented in this paper.




\section*{References}
\begin{thebibliography}{1}
\bibitem{Donoho}
Donoho D L 2006 {\it IEEE Trans. Inform. Theory} {\bf 52} 1289

\bibitem{CandesTao}
Cand\`{e}s E J and Tao T 2005 {\it IEEE Trans. Inform. Theory} {\bf 51} 4203

\bibitem{CRT}
Cand\`{e}s E J, Romberg J and Tao T 2006 {\it IEEE Trans. Inform. Theory} {\bf 52} 489

\bibitem{CSreview}
Cand\`{e}s E J and Wakin M B 2008 {\it IEEE Signal Process. Mag.} {\bf 25} 21

\bibitem{Elad}
Elad M 2010 {\it Sparse and redundant representations} (New York: Springer)

\bibitem{TW}
Tropp J A and Wright S J 2010 {\it Proc. of IEEE} {\bf 98} 948

\bibitem{DMM}
Donoho D L, Maleki A and Montanari A 2009
{\it Proc. Natl. Acad. Sci.} {\bf 106} 18914

\bibitem{Pearl}
Pearl J 1988 {\it Probabilistic reasoning in intelligent systems: networks
 of plausible inference} (San Francisco: Morgan Kaufmann)
 
\bibitem{DT}
Donoho D L and Tanner J 2005
{\it Proc. Natl. Acad. Soc.} {\bf 102} 9452

\bibitem{DMM3}
Donoho D L, Maleki A and Montanari A 2010
{\it Proc. of Information Theory Workshop, Cairo} ({\it Preprint} arXiv:0911.4219)

\bibitem{DMM4}
Donoho D L, Maleki A and Montanari A 2010
{\it Proc. of Information Theory Workshop, Cairo} ({\it Preprint} arXiv:0911.4222)

\bibitem{BM}
Bayati M and Montanari A 2011 {\it IEEE Trans. Inform. Theory}
 {\bf 57} 764

\bibitem{KWT}
Kabashima Y, Wadayama T and Tanaka T
 2009 {\it J. Stat. Mech.} L09003

\bibitem{lasso}
Tibshirani R 1996 {\it J. Royal Statist. Soc.} B  {\bf 58} 267

\bibitem{KW}
Kabashima Y and Wadayama T 2011 {\it Preprint} arXiv:1102.3220

\bibitem{KMSSZ}
Krzakala F, M\'{e}zard M, Sausset F, Sun Y and Zdeborov\'{a} L 2012 
{\it Phys. Rev.} X  {\bf 2} 021005

\bibitem{KMSSZ2}
Krzakala F, M\'{e}zard M, Sausset F, Sun Y and Zdeborov\'{a} L 2012 
{\it J. Stat Mech.} P08009

\bibitem{DT2}
Donoho D L and Tanner J 2009 {\it J. Amer. Math. Soc.} {\bf 22} 1

\bibitem{DT3}
Donoho D L and Tanner J 2009
{\it Phil. Trans. R. Soc.} A  {\bf 367} 4273

\bibitem{BLM}
Bayati M, Lelarge M and Montanari A 2012 {\it Proc. of 
IEEE International Symposium on Information Theory} 1643
({\it Preprint} arXiv:1207.7321)

\bibitem{TK}
Takeda K and Kabashima Y 2012 {\it Proc. of 46th Annual Conference 
on Information Sciences and Systems} ({\it Preprint} arXiv:1203.6246)

\end{thebibliography}
\end{document}